\documentclass[fleqn,usenatbib]{mnras}
\usepackage{newtxtext,newtxmath}
\usepackage[T1]{fontenc}
\usepackage{ae,aecompl}
\usepackage{multicol,multirow}
\usepackage{graphicx,stmaryrd}	
\usepackage{amsmath}	
\usepackage{amssymb}	

\title[Fragmentation around high vs. low mass stars]{Fragmentation favoured in discs around higher mass stars}

\author[J. Cadman et al.]{\parbox{\textwidth}
{James Cadman$^{1,2},$\thanks{E-mail: \texttt{cadman@roe.ac.uk}}
Ken Rice$^{1,2},$
Cassandra Hall$^{3},$
Thomas J. Haworth$^{4}$ \\
and Beth Biller$^{1,2}$}\vspace{0.4cm}
\\
$^{1}$SUPA, Institute for Astronomy, University of Edinburgh, Blackford Hill, Edinburgh, EH9 3HJ, Scotland, UK\\
$^{2}$Centre for Exoplanet Science, University of Edinburgh, Edinburgh, UK\\
$^{3}$Dept. of Physics and Astronomy, University of Leicester, University Road, Leicester, LE1 7RH, UK\\
$^{4}$Astronomy Unit, School of Physics and Astronomy, Queen Mary University of London, London, E1 4NS, UK\\
\\
}

\date{Accepted 2020 January 13. Received 2020 January 13; in original form 2019 November 19}

\pubyear{2020}

\begin{document}
\label{firstpage}
\maketitle

\begin{abstract}
We investigate how a protoplanetary disc's susceptibility to gravitational instabilities and fragmentation depends on the mass of its host star. We use 1D disc models in conjunction with 3D SPH simulations to determine the critical disc-to-star mass ratios at which discs become unstable against fragmentation, finding that discs become increasingly prone to the effects of self-gravity as we increase the host star mass. The actual limit for stability is sensitive to the disc temperature, so if the disc is optically thin stellar irradiation can dramatically stabilise discs against gravitational instability. However, even when this is the case we find that discs around $2$\,M$_{\odot}$ stars are prone to fragmentation, which will act to produce wide-orbit giant planets and brown dwarfs. The consequences of this work are two-fold: that low mass stars could in principle support high disc-to-star mass ratios, and that higher mass stars have discs that are more prone to fragmentation, which is qualitatively consistent with observations that favour high-mass wide-orbit planets around higher mass stars. We also find that the initial masses of these planets depends on the temperature in the disc at large radii, which itself depends on the level of stellar irradiation.
\end{abstract}




\section{Introduction}

Most protostellar discs will go through a self-gravitating phase during their lifetimes, most likely whilst they are still young and the disc is cold and massive \citep{linpringle87,linpringle90,riceetal10}. The susceptibility of a disc to the growth of a gravitational instability can be established by considering the Toomre parameter \citep{toomre64},
\begin{equation}
    Q = \frac{c_{\rm s} \kappa}{\pi G \Sigma}, 
\label{eq:Q}
\end{equation}
where $c_{\rm s}$ is the disc sound speed, $\kappa$ is the epicyclic frequency (equal to the angular frequency $\Omega$ in a rotationally supported disc), G is the gravitational constant, and $\Sigma$ is the disc surface density.

A self-gravitating phase will emerge when $Q \lesssim 1$. It can be seen from the dependence of $Q$ on $c_{\rm s}$ and $\Sigma$ why such a phase requires that the disc is sufficiently cool and/or massive.  The growth of the gravitational instability has two basic outcomes.  The disc will either settle into a long-lived \citep{halletal2019} quasi-steady state in which the instability acts to transport angular momentum \citep{paczynski78,laughlinbodenheimer94,lodatorice04}, or it can become sufficiently unstable that it fragments to form bound objects, potentially of planetary mass \citep{boss97, boss98}.\newline

A requirement for disc fragmentation is that the disc is able to cool rapidly \citep*{gammie01,ricelodatoarmitage05}.  In protostellar systems it is likely that these conditions will only be satisfied in the outer parts of the disc \citep{rafikov05,clarke09,ricearmitage09}, since small amounts of irradiation can act to suppress the instability \citep{halletal2016}. Models of the Jeans mass in spiral arms of self-gravitating discs predict that fragmentation will primarily form objects with initial masses greater than $\sim$3 Jupiter masses \citep{kratteretal10, forganrice11, forganrice13}. It is therefore likely that planet formation through the gravitational instability (hereafter GI) favours the formation of wide-orbit gas giants and brown dwarfs \citep{stamatellosetal09,forganrice13b,viganetal17,halletal2017,forganetal2018}.

Core accretion (hereafter CA) \citep{pollack96} describes planet formation through the steady collisional accumulation of smaller planetesimals to form progressively larger bodies. If these cores are able to grow massive enough, they may become capable of maintaining a massive gaseous envelope \citep{lissauer93, pollack96}, thus forming gas giant planets. However planetesimal growth in the outer disc is slow and planet formation timescales may well exceed disc lifetimes \citep*{haischladalada01}, making it challenging to explain the formation of wide-orbit gas giants through CA.

Results from radial velocity surveys for exoplanets suggest that giant planets are more frequently found around higher-mass hosts \citep{johnson07,bowler10}, although \citet{lloyd2011} express some concerns regarding how accurately the mass of these host stars can be measured. These results stimulated large-scale searches for directly imaged exoplanet companions around high-mass hosts \citep[primarily~A~stars,][]{janson11, vigan2012, nielsen2013}, even though intrinsically higher contrasts are needed to detect companions around these bright stars relative to solar analogues. Recently, considering the first 300 stars observed during the Gemini GPIES survey, \citet{nielsenetal19} found a significantly higher frequency of wide-orbit ($R=10-100$\,au) giant planets ($M=5-13$\,M$_{\rm Jup}$) around higher mass stars ($M>1.5$\,M$_{\odot}$) vs. $M<1.5$\,M$_{\odot}$ stars \citep{nielsenetal19}, while direct imaging surveys of low mass stars (M stars) have not yielded any companion detections \citep{lannieretal16}.  If wide-orbit giant planets are indeed preferentially formed via GI, these observations may suggest that fragmentation is favoured in discs around higher mass stars.

It has previously been shown that GI in discs around low-mass stars is quenched by a combination of viscous heating and stellar irradiation, making planet formation through fragmentation unlikely \citep*{matznerlevin05}. \cite{krattermatzner06} found a critical disc outer radius of $\sim 150$\,au, above which discs around massive stars may become prone to fragmentation.
This critical radius is set by two competing factors; increased stellar irradiation with increasing stellar mass pushing the radius out, while the more rapid accreting around the more massive stars favours fragmentation. \cite{kratterlodato16} used the scaling of $Q$ with disc-to-star mass ratio to suggest analytically that we may expect some scaling of instability with stellar mass. Recently this relation has been further explored by Haworth et al. (in prep.).

Haworth et al. (in prep.) demonstrated that low-mass stars are able to maintain discs with high disc-to-star mass ratios, with masses comparable to that of the central protostar, without becoming gravitationally unstable and fragmenting. The large mass reservoirs potentially available may have important consequences for planet formation through CA, and may help to explain the origin of multi-planet systems around very low-mass stars, such as Trappist-1 \citep{gillonetal17}. 

The work we present here is an extension of this previous work, but conversely aims to investigate how susceptibility to fragmentation varies with stellar mass. In particular we concentrate on the critical disc-to-star mass ratios for fragmentation. To approach this, 1D disc models for various stellar masses have been used to calculate the effective viscous-$\alpha$ values \citep{shakurasunyaev73,lodatorice04} for a range of disc radii and accretion rates. It has then been determined for which disc-to-star mass ratios we expect the disc to be unstable against fragmentation, assuming that disc fragmentation can occur when $\alpha \gtrsim 0.1$ \citep*{gammie01,ricelodatoarmitage05}. These 1D results have then been followed up using 3D smoothed particle hydrodynamics (SPH) simulations to validate their predictions.

This paper is organised as follows. In section \ref{sec:frag} we introduce fragmentation in self-gravitating discs and summarise previous work done in this area. In Sections \ref{sec:1Dsetup} and \ref{sec:3Dsetup} we describe the setup of our 1D disc models and 3D SPH simulations respectively, and present the results of these in section \ref{sec:results}. In section \ref{sec:jeansmass} we analyse the Jeans masses in self-gravitating discs, allowing us to predict the planet masses we might expect to form through disc fragmentation. In section \ref{sec:fragtimescale} we discuss the timescales over which we might expect the conditions for fragmentation to be satisfied. Finally, in sections \ref{sec:discuss} and \ref{sec:conclusion} we summarise our results and discuss their implications for planet formation through disc fragmentation. 

\section{Disc Fragmentation} \label{sec:frag}
As already mentioned, the stability of a rotating accretion disc against GI is characterised by the Toomre $Q$ parameter, shown in Equation (\ref{eq:Q}).  It is clear that the $Q$ parameter illustrates that GI is more likely in discs that are massive (large $\Sigma$) and/or cold (small $c_{\rm s}$).

A differentially-rotating disc is susceptible to axisymmetric perturbations if $Q < 1$ and to non-axisymmetric perturbations if $Q < 1.5 - 1.7$ \citep{durisenetal07}. In the latter case, the gravitational perturbations manifest themselves as spiral density waves, which can act as an effective means of transporting angular momentum.

If the disc settles into a quasi-steady state, in which heating and cooling are in balance, this can be parameterised via an effective viscosity with the effective viscous-$\alpha$ \citep{shakurasunyaev73} given by \citep{gammie01,ricelodatoarmitage05}
\begin{equation}\label{eq:alpha}
    \alpha = \frac{4}{9\gamma(\gamma-1)\beta_c},
\end{equation}
where $\gamma$ is the ratio of specific heats and $\beta_c = t_{\rm cool}\Omega$ is a dimensionless cooling parameter, with $t_{\rm cool}$ representing the local cooling timescale \citep{gammie01}. This modelling of disc viscosity assumes local angular momentum transport only, and may therefore be violated in some cases where global effects become important, as discussed in Section \ref{sec:1Dsetup}. A disc will typically be unstable against fragmentation if the local cooling time is smaller than the rate at which fragments are disrupted by the disc, given by the local dynamical time. Some early work \citep{gammie01, riceetal03} suggested that fragmentation occurs when $\beta_c \lesssim 3$. 

In an extension of this, \cite*{ricelodatoarmitage05} illustrated that the condition could be expressed in terms of a critical-$\alpha$ value, representing a maximum stress that a disc can sustain without fragmenting. Consistent with the results from \citet{gammie01}, they found $\alpha_{\rm crit} \approx 0.06$.  This, however, only really applies in the absence of an additional heating source.  The presence of an additional heating source, such as some kind of external irradiation, will tend to stabilise the disc both against fragmentation \citep{riceetal11} and the development of prominent spiral arms \citep{halletal2016, halletal2018}.  Fragmentation will then require more rapid cooling than in the absence of this additional heating source.  
There have, however, been suggestions \citep{merubate11,merubate12} that the simulations on which these estimates are based don't converge. In particular, the critical $\alpha$ decreases with increasing numerical resolution, suggesting that fragmentation could happen for very long cooling times. However, it now appears that this is more a consequence of numerical issues with the codes \citep{riceetal12,lodatoclarke11,paardekooperetal11, riceetal14,dengetal2017,hongpingetal19}, rather than an indication that fragmentation can actually happen for very long cooling times. More recent work continues to indicate that fragmentation typically requires $\alpha \approx 0.1$ \citep{baehretal17}.  We can't, however, rule out that there might be an element of stochasticity \citep{paardekooper12,youngclarke16,kleeetal17,kleeetal19}, or an alternative mode of fragmentation \citep{youngclarke15}, that could sometimes lead to fragmentation for longer cooling times, or smaller effective $\alpha$ values, than this boundary value. 

When a region of a gravitationally unstable disc fragments, it will collapse to form bound clumps of masses comparable to the local Jeans mass. In the presence of external irradiation this is given by, 

\begin{equation}\label{eq:jeansmass}
    M_J = \frac{\sqrt{3}}{32G}\frac{ \pi^3Q^{1/2}{c_{\rm s}}^2 H}{(1+4.47\sqrt{\alpha})^{1/2}}.
\end{equation}

Note that this expression differs slightly from the expression found previously in \citep{forganrice13}. We now have a different prefactor and the $1 + 4.47\sqrt{\alpha}$ term is now square-rooted. A full derivation of this expression can be found in Appendix \ref{sec:appendixA}.

Typical Jeans masses in the spiral arms of self-gravitating discs are found to be of the order a few Jupiter masses (see section \ref{sec:jeansmass}). We therefore expect, from the arguments presented in this section, that planet formation through fragmentation will primarily form wide-orbit giant planets, or brown dwarfs \citep{forganrice13}.

\section{1D Disc Models - Methodology}\label{sec:1Dsetup}

To investigate how disc stability against fragmentation varies with stellar mass, we have implemented the 1D disc models first presented by \citet{clarke09} and then further developed by \citet{forganrice11}.  Specifically, we use the formalism in which external irradiation is also included \citep{forganrice13}. We consider two cases; one in which irradiation leads to a constant background temperature of $T_{\rm irr}=10$\,K and another in which the stellar irradiation is based on the MIST stellar models for 0.5\,Myr stars \citep{MIST1, MIST2}. 

The four stellar masses considered in this analysis are $0.25$\,M$_{\odot}$, $0.5$\,M$_{\odot}$, $1.0$\,M$_{\odot}$ and $2.0$\,M$_{\odot}$. For host-star masses above $2$\,M$_\odot$, these models become complicated as the outer disc becomes optically thick and dynamical heating may become important. We therefore choose not to model stellar masses greater than this. For each stellar mass we have generated a suite of 1D disc models and investigated the conditions necessary for fragmentation to occur, assuming that fragmentation is possible for $\alpha \gtrsim 0.1$ \citep*{ricelodatoarmitage05}.

A self-gravitating disc is constructed by assuming that it settles into a state with a steady mass accretion rate given by \citep{pringle81}
\begin{equation}\label{eq:accretionrate}
    \dot{M} = \frac{3\pi\alpha c_{\rm s}^2 \Sigma}{\Omega}.
\end{equation}

Assuming that the disc is gravitationally unstable at all radii, with $Q = 2$, equations (\ref{eq:Q}), (\ref{eq:alpha}) and (\ref{eq:accretionrate}) allow for the values of $\alpha$, $\Sigma$ and $c_{\rm s}$ to be derived if we use that $\beta_c = (u/\dot{u})\Omega$ and that the cooling function is given by
\begin{equation}
    \dot{u} = \frac{\sigma_{SB}T^4}{\tau + 1/\tau},
\end{equation}
where $\sigma_{SB}$ is the Stefan-Boltzmann constant, $T$ is the disc temperature and $\tau$ is the optical depth. We can estimate the optical depth using $\tau = \Sigma\kappa(\rho,c_{\rm s})$, where $\rho=\Sigma/2H$ is the disc volume density and $\kappa$ is the opacity. Values of $\gamma$, $T$ and $\kappa$ are obtained from $\rho$ and $c_{\rm s}$ using the equation of state from \cite{stamatellosetal07}. The scaling of this cooling function with optical depth as $\tau + 1/\tau$ allows us to account for both optically thin and optically thick regimes. In regions of low optical depth, and regions of high optical depth, cooling will be inefficient and our cooling function will account for this.

In this way we have generated a suite of disc models for values of $\dot{M}$ and $R_{\rm out}$ in the ranges $10^{-10} - 10^{-1}$\,M$_{\odot}$\,yr$^{-1}$ and $1-200$\,au respectively. From the values of $\alpha$, $\Sigma$ and $c_{\rm s}$, we are able to calculate the disc-to-star mass ratio for each value of $\dot{M}$ and $R_{\rm out}$.

\begin{figure}
    \includegraphics[width=\linewidth]{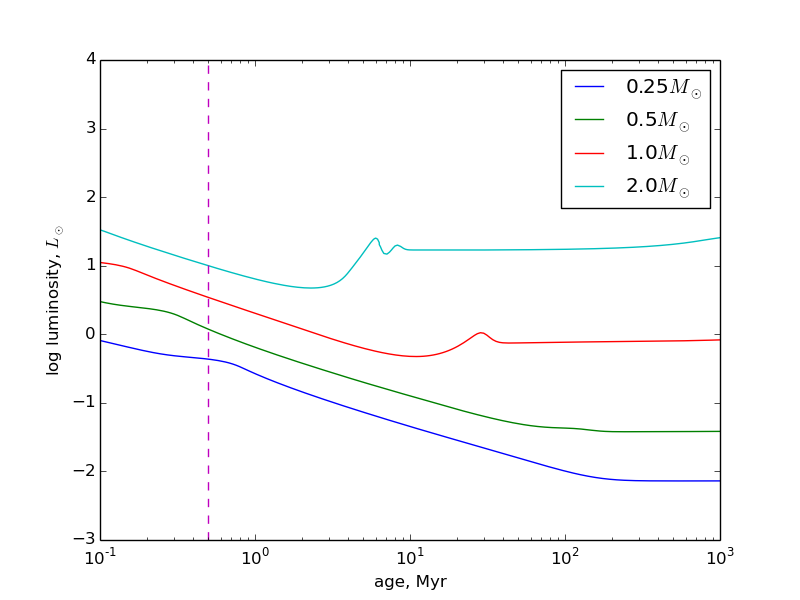}
    \caption{\label{fig:MISTtrakcs} MIST stellar evolution tracks \citep[see][]{MIST1, MIST2}. The 0.5Myr luminosities extracted from these plots have been used in these analyses.}
\end{figure}

For the case in which irradiation is modelled using a constant background temperature, the disc temperature is prevented from dropping below a floor of temperature of $T_{\rm irr}=10$ K. In the case of the stellar irradiated discs we model the temperature as,

\begin{equation}\label{eq:stellarirrad}
    T_{\rm irr} = \Big(\frac{L_*}{4\pi \sigma R^2}\Big)^{1/4},
\end{equation}
where $L_*$ is obtained from the MIST stellar evolution tracks at 0.5\,Myr \citep{MIST1,MIST2}. These tracks are plotted in Fig. \ref{fig:MISTtrakcs} and the values of $L_*$ used here for the cases of $0.25$\,M$_{\odot}$, $0.5$\,M$_{\odot}$, $1.0$\,M$_{\odot}$ and $2.0$\,M$_{\odot}$ stellar masses are $0.44$\,L$_{\odot}$, $1.19$\,L$_{\odot}$, $3.40$\,L$_{\odot}$ and $10.11$\,L$_{\odot}$ respectively. 

This modelling of stellar irradiation assumes the disc to be optically thin and thus passively irradiated. In reality there will be significant self-shielding in the inner disc and the true disc heating will lie somewhere in between these two cases. An initial assessment on the impact of self-shielding finds the mid-plane dust radiative equilibrium temperature to be a factor $\sim 3-4$ smaller than that from equation \ref{eq:stellarirrad}. We therefore might expect the true critical disc-to-star mass ratios to be closer to the predictions of the $T_{\rm irr}=10$\,K discs than the stellar irradiated discs (see Haworth et al. in prep. for a more detailed discussion).

Comparing the disc temperatures from Equation \ref{eq:stellarirrad} to the 10\,K irradiated discs we find that for a $R_{\rm out}=150$\,au disc in the cases of a $0.25$\,M$_{\odot}$, $0.5$\,M$_{\odot}$, $1$\,M$_{\odot}$ and $2$\,M$_{\odot}$ stellar host, the irradiation temperatures in the outer disc will be 26.1K, 33.5\,K, 43.6\,K and 57.2\,K respectively. These higher disc temperatures in the presence of stellar irradiation will further suppress fragmentation due to there being greater pressure support against gravitational collapse, whilst also reducing the disc effective-$\alpha$.

The 1D disc models presented here assume local angular momentum transport in which the disc viscosity can be represented by a local $\alpha-$parameter (e.g. Equation \ref{eq:alpha}). \cite{forganriceetal11} found the local approximation to be valid up to disc-to-star mass ratios of $q \sim 0.5$, above which global effects become important and the effective viscosity is not well represented by this local parameterisation. We should therefore proceed with caution when interpreting the results of these models at high disc-to-star mass ratios. However, they do provide useful information that informs the 3D SPH simulations which follow.

\section{3D SPH Simulations - Methodology}\label{sec:3Dsetup}

To extend the results from the 1D disc models we have produced a suite of 3D SPH simulations using the Phantom SPH code \citep{priceetal18}. We model the disc cooling using the radiative transfer method introduced in \cite{stamatellosetal07}. This method represents a more realistic cooling method than the simple $\beta-$cooling formalism, and allows us to consider regions of both high and low optical depths.

The gas discs are represented by 500,000 SPH particles, allowing us to simulate a large number of discs spanning a wide range of parameter space. The stellar masses are the same as those from the 1D models; $M_* = 0.25$\,M$_{\odot}$, $0.5$\,M$_{\odot}$, $1.0$\,M$_{\odot}$ and $2.0$\,M$_{\odot}$. Each disc has an initial surface density profile of $\Sigma \propto R^{-1.5}$, and an initial temperature profile, $T \propto R^{-0.5}$. These profiles were chosen to be consistent with those resulting from the 1D models. This steep surface density profile also avoids immediately inducing fragmentation artificially by initially having too much mass in the outer disc. In Haworth et al. (in prep.) shallower surface density and temperature profiles have been used, and we don't expect that these will affect our conclusions. We use artificial viscosity terms $\alpha_{\rm SPH}=0.1$ and $\beta_{\rm SPH}=0.2$. 

Once again we assume two cases of disc irradiation; one with a constant 10\,K floor temperature as well as stellar irradiation using the MIST 0.5\,Myr luminosities, in line with the 1D disc models.

For the cases of 10\,K and stellar irradiation, a total of 192 and 58 discs have been simulated respectively. The specific disc masses and radii were selected from inspection of the 1D model results, considering disc parameters which lie close to the $\alpha = 0.1$ contour. The inner disc radii are set as $R_{\rm in}=1.0$\,au with gas particles falling within this region being accreted onto the central protostar, represented here as a point mass.

Each disc has been allowed to evolve for 5 outer orbital periods, assuming that if it has not fragmented by this point then it will not fragment in the future. Discs are considered to have not fragmented if they initially appear to form clumps, but these clumps are then either rapidly destroyed by dynamical effects or accreted onto the central protostar within the 5 orbital periods.

\begin{figure*}
    \begin{multicols}{2}
        \includegraphics[width=\linewidth]{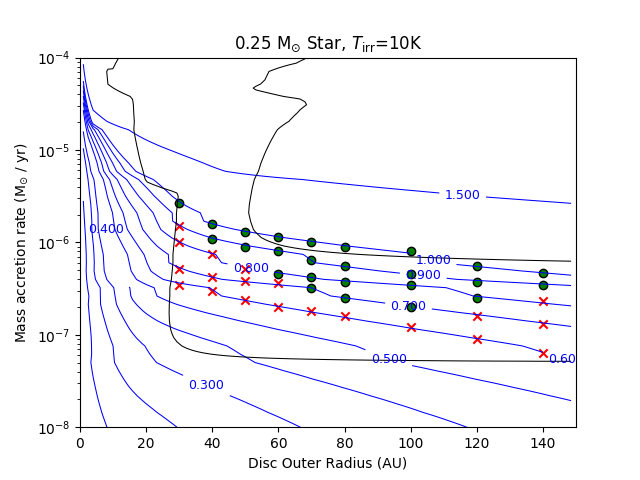}\par 
        \includegraphics[width=\linewidth]{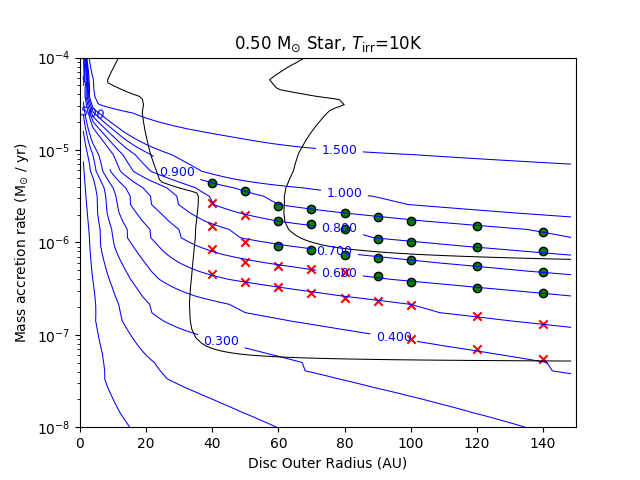}\par 
        \end{multicols}
    \begin{multicols}{2}
        \includegraphics[width=\linewidth]{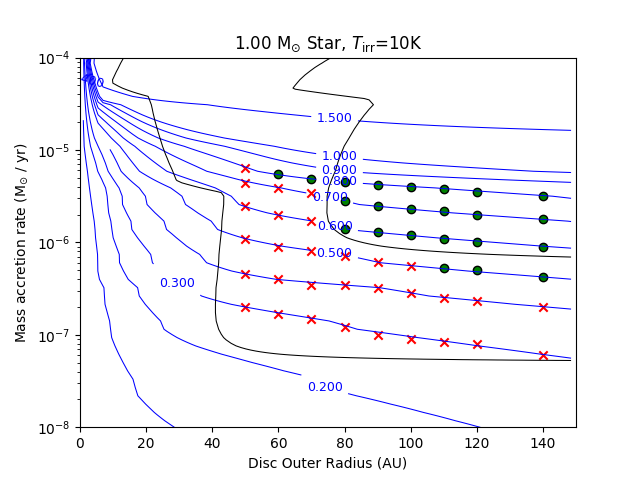}\par
        \includegraphics[width=\linewidth]{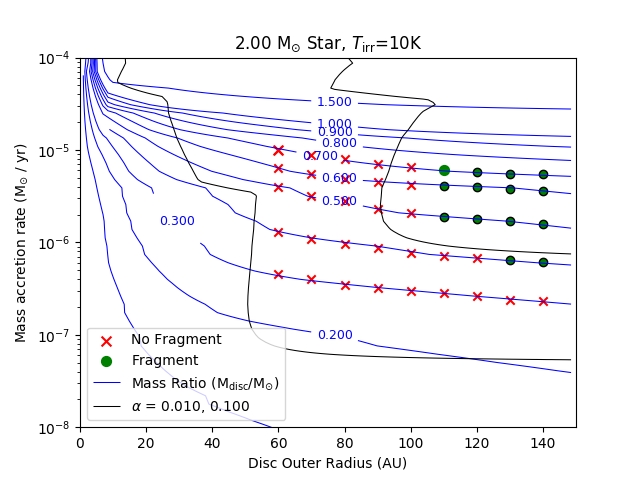}\par
    \end{multicols}

    \caption{\label{fig:1dmodels} Results of the 1D models (contours) and 3D SPH simulations (dots and crosses) for the case of 10\,K irradiated discs. The 2D contour plots show how the disc-to-star mass ratio (blue contours) varies as a function of accretion rate and disc outer radius for the cases of $0.25$\,M$_{\odot}$ (top left), $0.5$\,M$_{\odot}$ (top right), $1.0$\,M$_{\odot}$ (bottom left) and $2.0$\,M$_{\odot}$ (bottom right) host star masses. The results of the 3D SPH simulations are shown by the dots and crosses, representing fragmenting and non-fragmenting discs respectively. The effective Shakura-Sunyaev viscous-$\alpha$ is shown as black contours.}
\end{figure*}

\begin{figure*}
    \begin{multicols}{2}
        \includegraphics[width=\linewidth]{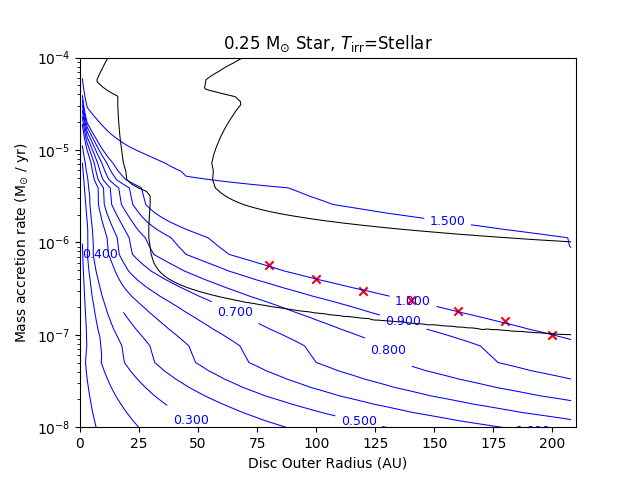}\par 
        \includegraphics[width=\linewidth]{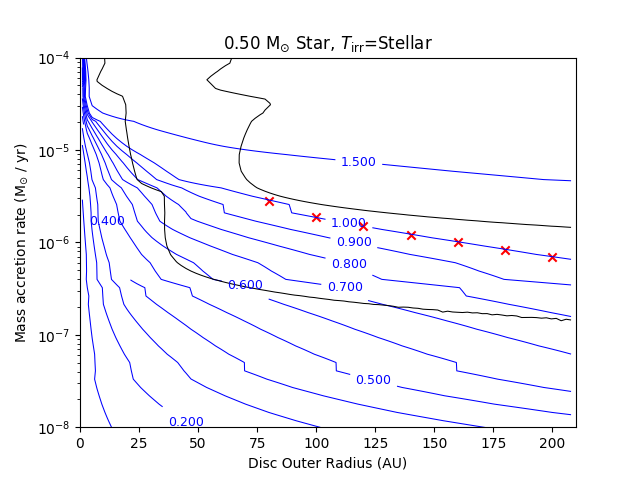}\par 
        \end{multicols}
    \begin{multicols}{2}
        \includegraphics[width=\linewidth]{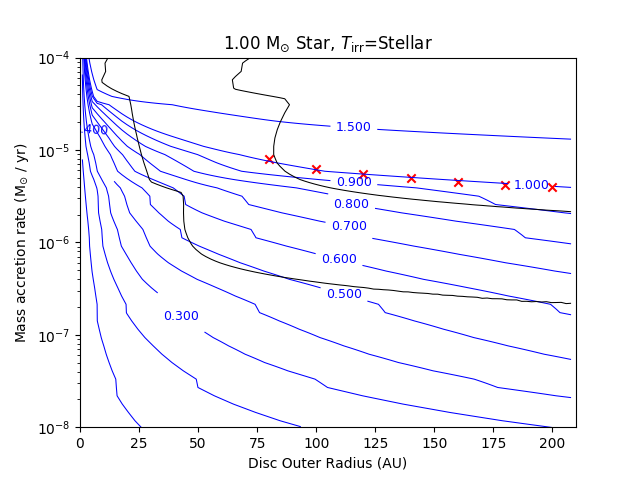}\par
        \includegraphics[width=\linewidth]{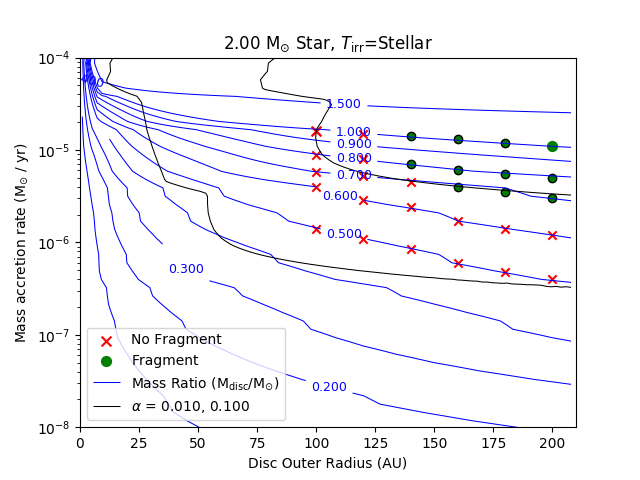}\par
    \end{multicols}

    \caption{\label{fig:1dmodelsstellar}  Results of the 1D models (contours) and 3D SPH simulations (dots and crosses) for the case of 0.5\,Myr MIST Stellar irradiated discs. The 2D contour plots show how the disc-to-star mass ratio (blue contours) varies as a function of accretion rate and disc outer radius for the cases of $0.25$\,M$_{\odot}$ (top left), $0.5$\,M$_{\odot}$ (top right), $1.0$\,M$_{\odot}$ (bottom left) and $2.0$\,M$_{\odot}$ (bottom right) host star masses. The results of the 3D SPH simulations are shown by the dots and crosses, representing fragmenting and non-fragmenting discs respectively. The effective Shakura-Sunyaev viscous-$\alpha$ is shown as black contours.}
\end{figure*}

\section{Results}\label{sec:results}

\subsection{1D Disc Models}

Considering initially Figures \ref{fig:1dmodels} and \ref{fig:1dmodelsstellar}, the blue contours show how the disc-to-star mass ratio, \textit{q}, varies with accretion rate as a function of disc outer radius.  For example, in Figure \ref{fig:1dmodels} for a $0.25$\,M$_{\odot}$ stellar host, a disc with an accretion rate $\dot{M} = 10^{-7}$\,M$_{\odot}$\,yr$^{-1}$ and a radius $R_{\rm out} = 80$\,au will have a disc-to-star mass ratio, $q=0.520$. 

The black contours show the Shakura-Sunyaev effective viscous-$\alpha$ values from Equation \ref{eq:alpha}. We show contours for $\alpha = 0.01$ and $\alpha = 0.1$.  As discussed in section \ref{sec:frag}, the canonical fragmentation boundary is typically taken to be $\alpha = 0.06$.  There is, however, some uncertainty in this exact value, partly due to convergence issues in the simulations \citep{merubate11}, partly due to possible stochasticity \citep{paardekooper12}, and partly because there is some evidence for an alternative mode of fragmentation \citep{youngclarke15}. It seems likely, though, that fragmentation will occur somewhere in the region between the $\alpha = 0.01$ and $\alpha = 0.1$ contours. What Figures \ref{fig:1dmodels} and \ref{fig:1dmodelsstellar} illustrate is that this will require discs with masses that are a significant fraction of the mass of the central protostar.  

\subsubsection{$T_{\rm irr}=10$\,K}

Figure \ref{fig:1dmodels} shows the scenario in which we assume some background irradiation prevents the disc temperature from dropping below $T = 10$\,K.  It shows that as we increase the host star mass from $0.25$\,M$_{\odot}$ to $2$\,M$_{\odot}$ the critical mass ratio for the discs to become unstable against fragmentation generally decreases. If we consider the $\alpha = 0.1$ contour in Figure \ref{fig:1dmodels}, for a $0.25$\,M$_{\odot}$ stellar host the disc-to-star mass ratio needs to exceed $q=1$ before the disc's viscous-$\alpha$ values exceed $\alpha = 0.1$. We would therefore expect these discs to avoid fragmenting even for very large disc-to-star mass ratios.  As stellar mass is increased to $2$\,M$_{\odot}$, the disc's viscous$-\alpha$ exceeds $\alpha = 0.1$ for mass ratios of around $q=0.4-0.5$. The minimum radius for fragmentation also tends to shift outwards with increasing stellar mass. Fragmentation is only expected in discs larger than $R\sim 90$\,au in the case of a $2$\,M$_{\odot}$ stellar host, compared to $R\sim 50$\,au in the case of a $0.25$\,M$_{\odot}$ stellar host.

\subsubsection{$T_{\rm irr}=\rm Stellar$}

When considering the case of stellar irradiation, shown in Figure \ref{fig:1dmodelsstellar}, the critical mass ratios are now shifted to even higher masses with respect to the case when $T_{\rm irr}=10$\,K. This is due to the now higher disc temperatures suppressing GI \citep{riceetal11}.  For a $0.25$\,M$_{\odot}$ stellar host we now require $q \gtrsim 1.4$ before the disc's viscous-$\alpha$ values exceed $\alpha = 0.1$. Increasing the stellar mass to $2$\,M$_{\odot}$ reduces the required disc-to-star mass ratio to $q \gtrsim 0.7$. The minimum radii at which fragmentation is likely to occur has also been pushed outward with respect to the 10\,K irradiated discs. Fragmentation will now only occur in discs larger than $R\sim 100$\,au for a $2$\,M$_{\odot}$ stellar host, and $R\sim 60$\,au for a $0.25$\,M$_{\odot}$ stellar host.
\\
\\
These 1D models therefore indicate that fragmentation requires higher disc-to-star mass ratios around lower mass stars than around higher mass stars. When including the effects of stellar irradiation we also find that discs become less prone to fragmentation, as we now require far higher disc-to-star mass ratios before the discs' viscous$-\alpha$ values exceed $\alpha = 0.1$. Hence, these 1D models suggest that disc fragmentation may favour higher mass stellar hosts. We note again that above $q\sim0.5$, global effects may become important which are not accounted for in these 1D models.  However, we do not expect this to affect the general trends demonstrated by these results.

\begin{figure*}
    \includegraphics[width=\linewidth]{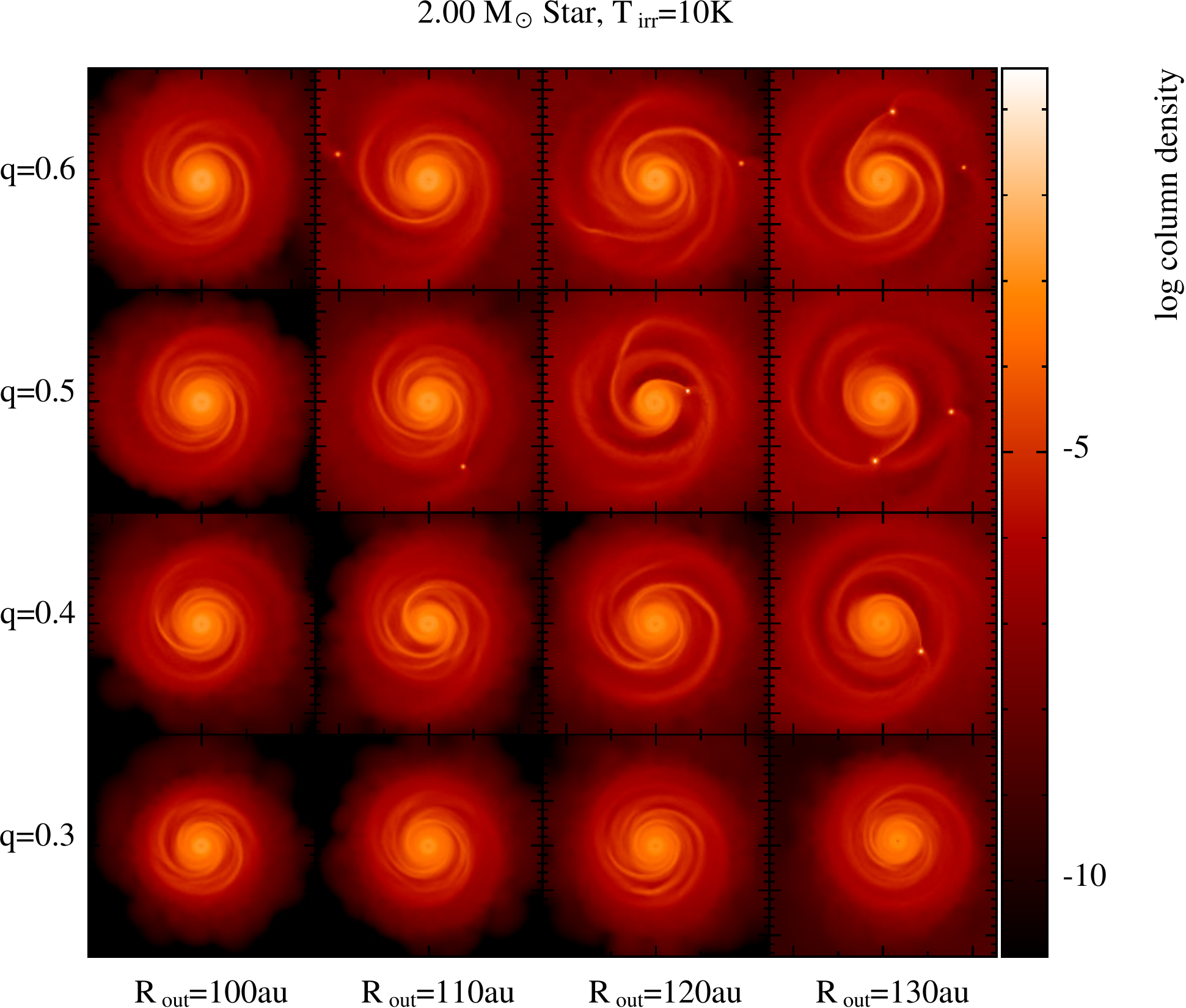}\par 
    \caption{\label{fig:3dplots2solarmass}  3D SPH results demonstrating how discs become more gravitationally unstable and prone to fragmentation as we increase the disc-to-star mass ratio and the disc outer radius. The discs shown here are for a $2$\,M$_{\odot}$ host star and $T_{\rm irr} = 10$\,K. Each disc has been allowed to evolve for 5 outer orbital periods, with only the largest and most massive discs having formed bound fragments.}
\end{figure*}
\begin{figure*}
    \includegraphics[width=\linewidth]{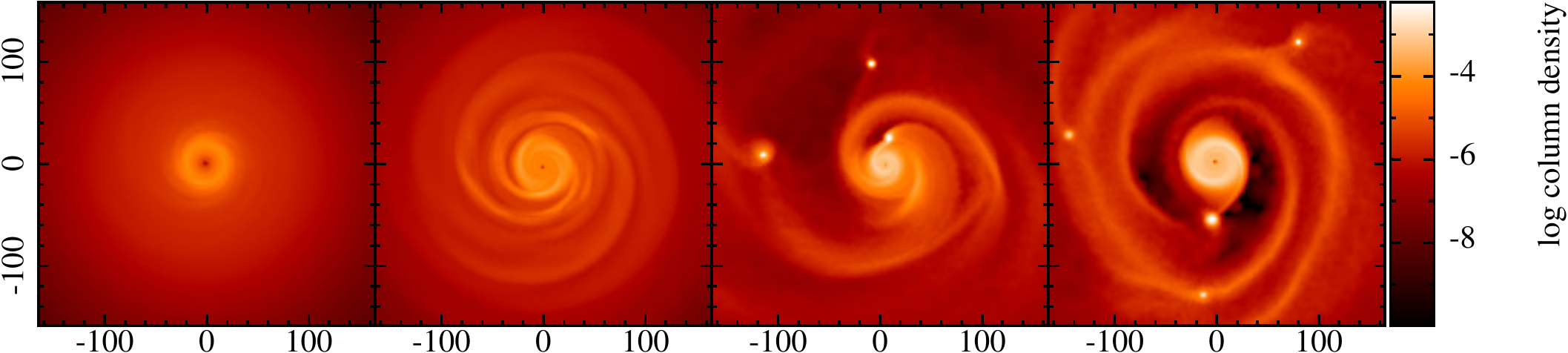}\par 
    \caption{\label{fig:10KSPHdiscs} 3D SPH results showing how the final states of the discs vary with stellar mass in the case of $T_{\rm irr}=10$\,K. The discs shown have mass ratios, $q=0.5$, and outer radii, $R_{\rm out}=140$\,au, with stellar masses, from left to right, of $0.25$\,M$_{\odot}$, $0.5$\,M$_{\odot}$, $1.0$\,M$_{\odot}$ and $2.0$\,M$_{\odot}$.}
\end{figure*}
\begin{figure*}
    \includegraphics[width=\linewidth]{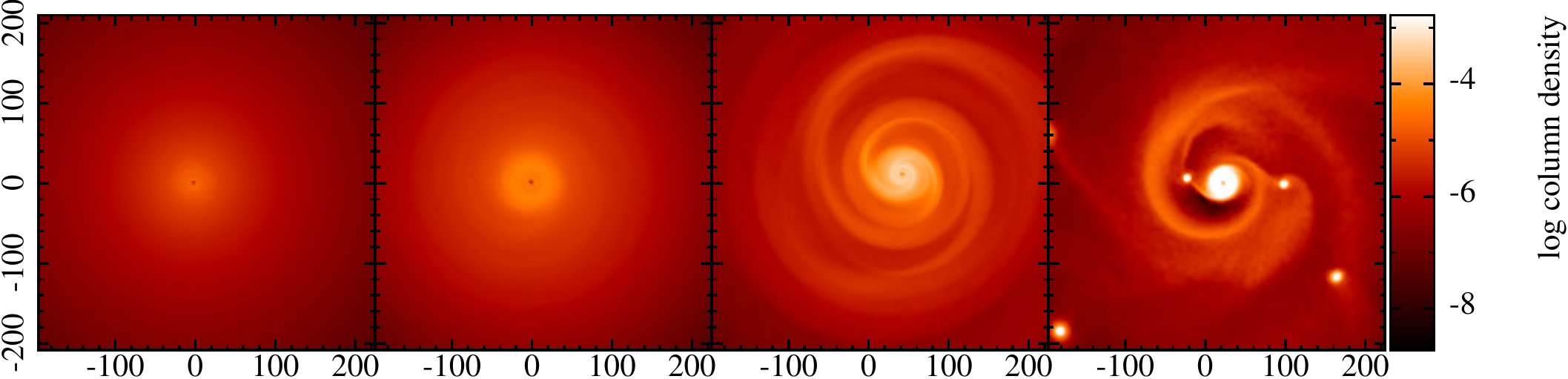}\par
    \caption{\label{fig:stellarSPHdiscs} Results of the 3D SPH simulations in the case of 0.5\,Myr MIST Stellar irradiated discs. The discs shown have mass ratios, $q=1.0$, and outer radii, $R_{\rm out}=200$\,au, with stellar masses, from left to right, of $M_* = 0.25$\,M$_{\odot}$, $0.5$\,M$_{\odot}$, $1.0$\,M$_{\odot}$ and $2.0$\,M$_{\odot}$.}
\end{figure*}

\subsection{3D SPH Simulations}\label{sec:3Dresults}

In Figures \ref{fig:1dmodels} and \ref{fig:1dmodelsstellar}, we also show the results of the 3D SPH simulations. These are represented by the markers over-plotted on the mass-ratio contours. Each marker represents an individual simulation, which has been set up as described in Section \ref{sec:3Dsetup}. Red crosses show discs that have not fragmented after 5 outer orbital periods and green circles show discs in which a bound fragment has formed.

Example plots of the final states of these simulated discs are displayed in Figure. \ref{fig:3dplots2solarmass}. The discs shown are for a $2$\,M$_{\odot}$ host star in the case of $T_{\rm irr}=10$\,K, and demonstrate how discs become increasingly gravitationally unstable and prone to fragmentation as we increase the disc's outer radius and mass. Bound fragments have clearly formed in the largest and most massive discs, whilst the smaller and less massive discs display spiral arm structure only. 

As we mentioned previously, our 1D models assume local angular momentum transport which may not be valid at high disc-to-star mass ratios. The effect of this can be clearly seen in Figure \ref{fig:1dmodels} by comparing the 1D predictions to the 3D results at the highest mass ratios ($q \geq 0.5$) in e.g. the $0.25$\,M$_{\odot}$ case. Here we find that fragmentation can occur for lower $q$ values than initially predicted from the 1D models. When comparing the 1D predictions to the 3D results for slightly lower mass ratios, e.g. from the $2.0$\,M$_{\odot}$ results, we find the results to be far more consistent as the 1D models are now more reliable. Despite this, the SPH results shown in Figures \ref{fig:1dmodels} and \ref{fig:1dmodelsstellar} display the same general trend as suggested by the 1D disc models; the critical disc-to-star mass ratio for fragmentation generally decreases with increasing stellar host mass and the critical radius steadily shifts outward.

In Figure \ref{fig:1dmodels}, for a $0.25$\,M$_{\odot}$ host star and $T_{\rm irr}=10$\,K, discs are able to fragment for mass ratios greater than $q=0.7$. This is lower than suggested by the $\alpha=0.1$ contour but still broadly consistent with the 1D models. For its $2$\,M$_{\odot}$ counterpart, discs are able to fragment for mass ratios of $q=0.4$ and above. Discs as small as $R_{\rm out}=30$\,au are able to fragment around a $0.25$\,M$_{\odot}$ host star, with this value increasing to $R_{\rm out}=110$\,au for a $2$\,M$_{\odot}$ stellar host.

In Figure \ref{fig:1dmodelsstellar}, when considering stellar irradiated discs, only the $2$\,M$_{\odot}$ stellar hosts produce fragments after 5 orbital periods. Fragmentation can occur for $q \geq 0.7$ in these systems. All other stellar hosts had no fragmentation for discs with $q=1.0$, with these being the highest mass discs modelled in our SPH simulations. We have chosen to not model discs with mass ratios greater than this as it is unclear whether these would exist as disc-star systems at all, or whether the system would instead be deeply embedded in an envelope. See Section \ref{sec:discusslowmass} for a discussion of the implications of high disc-to-star mass ratios. The critical radii at which we expect discs to fragment has again shifted outward with respect to the $10$\,K irradiated discs, with discs around a $2$\,M$_{\odot}$ star only fragmenting for $R_{\rm out}\gtrsim 140$\,au.

Figures \ref{fig:10KSPHdiscs} and \ref{fig:stellarSPHdiscs} further illustrate the effects of increasing stellar host mass on disc instability. It can be seen that as we increase the star mass from left to right for constant $R_{\rm out}$ and $q$, discs become increasingly gravitationally unstable. In Figure \ref{fig:10KSPHdiscs} in the case of $T_{\rm irr}=10$\,K, $q=0.5$ and $R_{\rm out}=140$\,au, the discs with a $1$\,M$_{\odot}$ and a $2$\,M$_{\odot}$ host star have formed bound fragments, whilst for a $0.5$\,M$_{\odot}$ host star we observe spiral structure and for a $0.25$\,M$_{\odot}$ host star we observe almost no spiral structure at all. A similar trend can be seen for the case of stellar irradiation in Figure \ref{fig:stellarSPHdiscs}, with only the $2$\,M$_{\odot}$ host star case forming bound fragments.

The results from these 3D models are therefore largely consistent with the 1D calculations, suggesting that fragmentation is preferred at large radii around higher mass stellar hosts. For an optically thin disc we find that stellar irradiation is potentially able to completely switch off fragmentation around lower mass stars. However we again note that the disc is likely to be optically thick in the inner regions, and that the true critical $q$ values may lie closer to the $T_{\rm irr}=10$\,K case.

\begin{figure*}
    \begin{multicols}{2}
        \includegraphics[width=\linewidth]{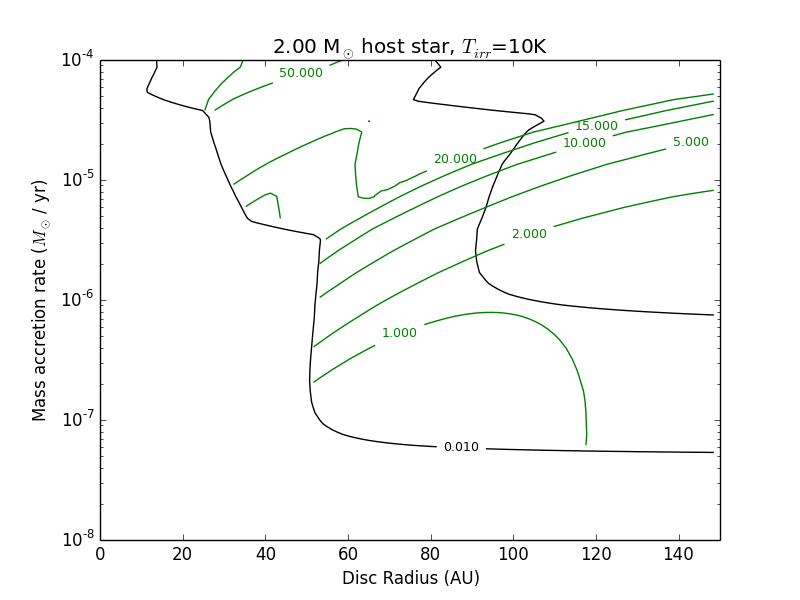}\par 
        \includegraphics[width=\linewidth]{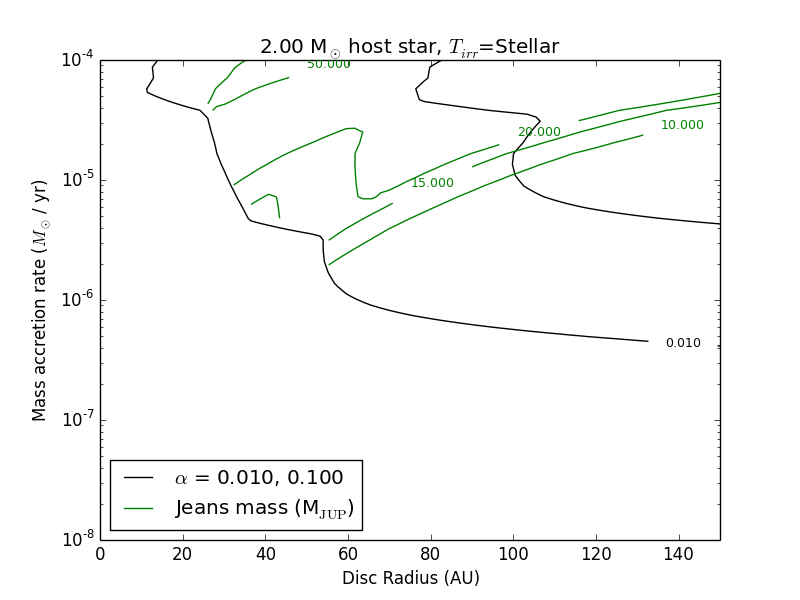}\par 
    \end{multicols}
    \caption{\label{fig:jeansmasses} Predicted Jeans masses from Equation \ref{eq:jeansmass} in discs around a $2$\,M$_{\odot}$ host star for the cases of $T_{\rm irr}=10$\,K (left) and stellar irradiation (right).}
\end{figure*}

\section{A reevaluation of the Jeans mass in a spiral wave perturbation}\label{sec:jeansmass}

As Equation \ref{eq:jeansmass} differs from that found in \cite{forganrice13}, with a full derivation found in Appendix \ref{sec:appendixA}, it is necessary to re-analyse the Jeans masses inside spiral density perturbations here. The Jeans mass in the spiral arms of self-gravitating discs represents the masses of the fragments that we expect to form in these regions, thus placing a constraint on the type of objects which may be produced through disc fragmentation. 

Figure \ref{fig:jeansmasses} shows the calculated Jeans masses from the 1D disc models in the case of a $2$\,M$_{\odot}$ host star, considering both $10$\,K and stellar irradiation. We consider here the case of a $2$\,M$_{\odot}$ host star as we are concerned with fragmentation around the more massive stellar hosts. The Jeans mass for each value of $\dot{M}$ and $R_{\rm out}$ has been calculated using Equation \ref{eq:jeansmass} and plotted as the    
green contours in Figure \ref{fig:jeansmasses}.

The minimum Jeans masses for the $10$\,K and stellar irradiated cases are $1.10$\,M$_{\rm Jup}$ and $6.18$\,M$_{\rm Jup}$ respectively, assuming fragmentation is only possible above the $\alpha=0.1$ contour. The tendency for the Jeans mass to increase with the level of irradiation is a consequence of higher disc temperatures reducing the effective-$\alpha$ thus causing discs to be more massive for a given $\dot{M}$ and $R_{\rm out}$, and the higher temperatures producing greater pressure support against gravitational collapse, as previously discussed in \cite{forganrice13}.

The analysis in \cite{forganrice13} considered the case of a $1$\,M$_{\odot}$ stellar host, and it should be noted that the values found here remain reasonably similar to those found previously despite the changes made to Equation \ref{eq:jeansmass}. For the case of a $1$\,M$_{\odot}$ stellar host, we find minimum Jeans masses of $1.10$\,M$_{\rm Jup}$ and $4.60$\,M$_{\rm Jup}$ when using 10\,K and stellar irradiation respectively, compared to values of $4.1$\,M$_{\rm Jup}$ and $11.2$\,M$_{\rm Jup}$ found in \cite{forganrice13} previously.

\section{Timescale for fragmentation}
\label{sec:fragtimescale}
Our results indicate that fragmentation is preferred in discs around higher mass stars, and could potentially be completely suppressed in very-low-mass stars if the level of irradiation is sufficient. However, another factor to consider is the timescale over which a disc may sustain the conditions that are suitable for fragmentation. This is not possible to assess using the results 
from the 1D model and the 3D SPH simulations, since the 1D models are not time-dependent and the 3D SPH simulations are simply sampling regions of parameter space.

To consider this, we use the time-dependent models presented in \citet{ricearmitage09}, which assume
that angular momentum transport is predominantly driven by GI. Given that we don't actually know what
the initial conditions will be, we assume that all discs start with an outer radius of $R_{\rm out} = 100$ au and with a 
disc-to-star mass ratio of $q = 1$. We also only consider the case where $T_{\rm irr} = 10$ K. 

\begin{figure}
    \centering
    \includegraphics[width=\linewidth]{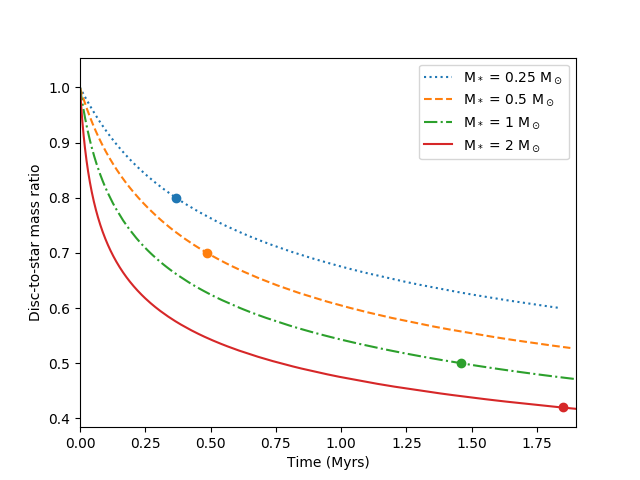}
    \caption{Figure showing the evolution of disc-to-star mass ratio, $q$, in discs in which the gravitational instability is the dominant angular momentum transport mechanism, for host star masses of $M_* = 0.25, 0.5, 1$ and $2$ M$_\odot$. The markers show the disc-to-star mass ratios above which disc fragmentation is possible, based on the results presented in Figure \ref{fig:1dmodels}.}
    \label{fig:time-dependent}
\end{figure}

Figure \ref{fig:time-dependent} shows the time evolution of the disc-to-star mass ratio for the same host star masses as considered before ($M_* = 0.25, 0.5, 1$ and $2$ M$_\odot$). The markers show, for each host star, the disc-to-star mass ratio above which fragmentation is possible, based on the results presented in Figure \ref{fig:1dmodels}.  
What Figure \ref{fig:time-dependent} illustrates is that, in conjunction with the required disc-to-star mass ratio decreasing with increasing stellar mass, the timescale over which fragmentation could occur also increases.  

Of course, Figure \ref{fig:time-dependent} does assume that sufficiently massive discs can indeed exist, but - if they can - the conditions for fragmentation would only persist around a 0.25 M$_\odot$ host star for a few 100 kyr.  Around a 2 M$_\odot$ host star, however, the timescale for fragmentation could be much longer, potentially a Myr, or longer.  However, this does assume that GI is the dominant mass transport mechanism, which may not be the case once the disc mass, and mass accretion rate, have become low enough for other mechanisms to become more important \citep{riceetal10}.

\section{Discussion}\label{sec:discuss}

\subsection{Implications for planet formation via disc fragmentation}\label{sec:discusslowmass}

The results presented in Section \ref{sec:results} illustrate that disc fragmentation is potentially favoured around higher-mass stars.  If we consider the case where $T_{\rm irr} = 10$\,K, and assume that the fragmentation boundary is at $\alpha = 0.1$, fragmentation requires a disc-to-star mass ratio of close to unity for a $0.25$\,M$_{\odot}$ host star, but requires $q \sim 0.4$ around a $2$\,M$_{\odot}$ host.  If we then consider stellar irradiation (e.g. Figure \ref{fig:1dmodelsstellar}), fragmentation around a $0.25$\,M$_{\odot}$ host star would then require disc masses that exceed the mass of the central protostar, while fragmentation around a $2$\,M$_{\odot}$ host could still occur for mass ratios of $q \sim 0.6$. This might suggest that stellar irradiation could completely suppress fragmentation around lower-mass host stars.

However, the simple modelling of stellar irradiation used in these models does not account for these being young, massive discs and there likely being a large amount of material in the inner disc regions. We therefore neglect factors such as self-shielding by material in the inner disc that could lead to stellar irradiation having less of an impact at large radii than we've assumed here. We should therefore expect that the true heating to be somewhere in between the two irradiation cases we've considered, (possibly being closer to the $T_{\rm irr}=10$\,K as already mentioned in section \ref{sec:1Dsetup}) and that the critical mass ratio where fragmentation can occur is probably somewhere within the range we've presented. We don't expect, however, that this will influence the trend that fragmentation is preferred around higher-mass host stars.

In general, discs with mass ratios of order unity, or above, are probably unrealistic. For Class II sources we expect the disc mass to be small compared to the stellar mass, usually no more than 10$\%$. Higher mass-ratio systems would likely be in the Class I phase whilst there is still a large amount of material in the envelope. For even higher mass ratio systems, with $q$ approaching unity, we would expect them to likely still be in the Class 0 phase in which the source is still deeply embedded. In this phase it is uncertain if there would be a star-disc system at all, or if instead there would be a massive envelope or torus. Additionally, even if such a system could exist, it would probably evolve very rapidly.  It's, therefore, unclear if there would be sufficient time for fragmentation to actually occur in a disc with $q > 1$.

A full discussion for the implications of lower mass stars being capable of hosting high-mass-ratio discs before becoming susceptible to gravitational instabilities can be found in Haworth et al. (in prep.). The key points to note are that the results suggest these systems may potentially have very large mass reservoirs available to them for planet formation through CA, thus loosening the constraint that any formation scenario (e.g., the Trappist-1 system, \citealt{gillonetal17}) must involve highly efficient dust growth. Haworth et al. (in prep.) also find that the high mass ratio discs ($q\gtrsim0.3$) required from photoevaporation models of the formation of the Trappist-1 system \citep{haworthetal16} to be entirely plausible, with our models finding discs to be gravitationally stable even when far more massive than this.

It is then intriguing that \cite{moralesetal19} recently discovered a $0.46$\,M$_{\rm Jup}$ planet orbiting a very low mass, $0.12$\,M$_{\odot}$, M dwarf on a 204 day period, with the authors proposing GI as the likely formation scenario. The results presented here suggest that only very massive discs around these very low mass stars may be permitted to be gravitationally unstable, thus indicating that such massive discs may indeed exist. We also require that these discs be optically thick to stellar irradiation, which would likely be the case for such a massive disc.

The results in this paper are complementary to those presented in Haworth et al. (in prep.). Fragmentation around lower mass stars requires large disc-to-star mass ratios and could be completely suppressed in the presence of stellar irradiation.  However, the required disc-to-star mass ratio decreases with increasing central star mass and the mass ratio required for fragmentation remains below $q \sim 1$ for higher mass stars, even in the presence of stellar irradiation. Our results therefore find fragmentation to be preferred around higher-mass stars ($M_* \sim 2$\,M$_{\odot}$) and an unlikely, if not an altogether impossible, planet formation scenario around very-low-mass stars.

Several direct imaging surveys for companions around $M>1.5$\,M$_{\odot}$ stars have tentatively pointed to a higher fraction of exoplanet and brown dwarf companions to higher mass stars relative to solar analogues or very-low-mass stars \citep{janson11, nielsen2013, vigan2012}. Recently, considering the first 300 stars observed during the Gemini GPIES survey, \citet{nielsenetal19} demonstrates this more conclusively, finding a significantly higher frequency of wide-orbit ($R=10-100$\,au) giant planets ($M=5-13$\,M$_{\rm Jup}$) around higher mass stars ($M>1.5$\,M$_{\odot}$) vs. $M<1.5$\,M$_{\odot}$ stars.  \citet{nielsenetal19} find an occurrence rate of (10-100 au) giant planets ($M=5-13$\,M$_{\rm Jup}$) of $9^{+5}_{-4} \%$ for their high mass stellar sample, vs. a brown dwarf occurrence rate ($M=13-80$\,M$_{\rm Jup}$, 10-100 AU) of $0.8^{+0.8}_{-0.5} \%$ around all survey stars.  The mass divisions adopted in \citep{nielsenetal19} do not straightforwardly map to a specific formation mechanism -- the brown dwarfs they detect could likely have formed via gravitational instability, whereas some of the planets in their cohort (e.g. 51 Eri b, for instance) are likely lower than the Jeans masses we have calculated here, and thus not as likely to be disc instability objects.  However, these results imply that the total companion frequency ($M=5-80$\,M$_{\rm Jup}$, 10-100 AU) must be higher for their high mass vs. low mass stellar sample, qualitatively consistent with the work presented here.

Although our analysis does indicate that disc fragmentation is more likely around higher-mass stars, it also suggests that it will probably occur at radii $\gtrsim 100$\,au, with this critical radius moving outward with increased levels of irradiation.  We might also expect the fragments to have initial masses above $5$\,M$_{\rm Jup}$ and to undergo further growth.  However, since we expect GI to act when the disc is young ($< 0.1$\,Myr) and massive, we would expect these fragments to undergo significant inward radial migration and potentially tidal downsizing after they form \citep[e.g. see][]{nayakshin10,forganrice13}. 

\section{Conclusions}\label{sec:conclusion}

In this paper we have used a set of 1D disc models followed by a suite of 3D SPH simulations to investigate how the conditions necessary for gravitational instability in protostellar discs vary with host star mass. In these models we have varied the disc masses and radii and have in particular focused on determining the critical disc-to-star mass ratio at which fragmentation is able to occur for stellar masses $M_* = 0.25$\,M$_{\odot}$, $0.5$\,M$_{\odot}$, $1$\,M$_{\odot}$ and $2$\,M$_{\odot}$. We have run models for both $T_{\rm irr}=10$\,K and for stellar irradiation, with the true disc irradiation likely lying somewhere in between these two cases.

The primary conclusions drawn from this work are that,
\begin{enumerate}
    \item Discs become more susceptible to GI as we increase the host star mass, with discs around higher mass stars being prone to fragmentation which will tend to produce wide-orbit giant planets and brown dwarfs. 
    \item Discs around lower mass stars ($M \leq 1.0$\,M$_{\odot}$) are able to host very high mass-ratio discs whilst still remaining gravitationally stable. In the case of stellar irradiated discs, when using the 0.5\,Myr MIST stellar luminosities in the optically thin regime, we find fragmentation to be completely suppressed in discs up to mass ratios of order unity. This may have important implications for CA, since it may be possible for these discs to have large mass reservoirs available for planet formation.  This could allow for less strict constraints with regards to pebble accretion efficiency and the depletion of disc material in planet formation models.
    \item Discs around higher-mass stars $M \geq 2$\,M$_{\odot}$ are more susceptible to GI and fragmentation. For the case of a $2$\,M$_{\odot}$ host star, we find that discs may fragment for mass ratios $q\geq0.4$ and $q\geq0.7$ in the cases of $T_{\rm irr}=10$\,K and 0.5\,Myr MIST stellar irradiated discs respectively. We find that fragmentation will only likely occur at radii $\gtrsim 100$\,au, with this critical radius increasing with increased irradiation and with increasing host star mass. Fragment masses are found to be strongly dependent on disc irradiation, with hotter discs producing more massive planets due to higher Jeans masses. Fragmentation in discs around $2$\,M$_{\odot}$ stars will produce objects of masses $\geq 1.10$\,M$_{\rm Jup}$ and $\geq 6.18$\,M$_{\rm Jup}$ in discs with $T_{\rm irr}=10$\,K and stellar irradiation respectively, thus producing wide orbit giant planets and brown dwarfs.
    \item Discs around $2$\,M$_{\odot}$ stars are able to sustain the conditions necessary for fragmentation for far longer timescales than discs around lower mass stars are. This is due to these discs becoming unstable against fragmentation for lower disc-to-star mass ratios, thus the conditions necessary for discs to be unstable against fragmentation will be satisfied for longer.
\end{enumerate}

\section*{Acknowledgements}

CH is a Winton Fellow and this research has been supported by Winton Philanthropies / The David and Claudia Harding Foundation. TJH is funded by a Royal Society Dorothy Hodgkin Fellowship.



\bibliographystyle{mnras}
\bibliography{main}

\begin{thebibliography}{}
\makeatletter
\relax
\def\mn@urlcharsother{\let\do\@makeother \do\$\do\&\do\#\do\^\do\_\do\%\do\~}
\def\mn@doi{\begingroup\mn@urlcharsother \@ifnextchar [ {\mn@doi@}
  {\mn@doi@[]}}
\def\mn@doi@[#1]#2{\def\@tempa{#1}\ifx\@tempa\@empty \href
  {http://dx.doi.org/#2} {doi:#2}\else \href {http://dx.doi.org/#2} {#1}\fi
  \endgroup}
\def\mn@eprint#1#2{\mn@eprint@#1:#2::\@nil}
\def\mn@eprint@arXiv#1{\href {http://arxiv.org/abs/#1} {{\tt arXiv:#1}}}
\def\mn@eprint@dblp#1{\href {http://dblp.uni-trier.de/rec/bibtex/#1.xml}
  {dblp:#1}}
\def\mn@eprint@#1:#2:#3:#4\@nil{\def\@tempa {#1}\def\@tempb {#2}\def\@tempc
  {#3}\ifx \@tempc \@empty \let \@tempc \@tempb \let \@tempb \@tempa \fi \ifx
  \@tempb \@empty \def\@tempb {arXiv}\fi \@ifundefined
  {mn@eprint@\@tempb}{\@tempb:\@tempc}{\expandafter \expandafter \csname
  mn@eprint@\@tempb\endcsname \expandafter{\@tempc}}}

\bibitem[\protect\citeauthoryear{{Baehr}, {Klahr}  \& {Kratter}}{{Baehr}
  et~al.}{2017}]{baehretal17}
{Baehr} H.,  {Klahr} H.,   {Kratter} K.~M.,  2017, \mn@doi [\apj]
  {10.3847/1538-4357/aa8a66}, \href
  {https://ui.adsabs.harvard.edu/abs/2017ApJ...848...40B} {848, 40}

\bibitem[\protect\citeauthoryear{Boss}{Boss}{1997}]{boss97}
Boss A.~P.,  1997, Science, 276, 1836

\bibitem[\protect\citeauthoryear{Boss}{Boss}{1998}]{boss98}
Boss A.~P.,  1998, Nature, 393, 141

\bibitem[\protect\citeauthoryear{{Bowler} et~al.,}{{Bowler}
  et~al.}{2010}]{bowler10}
{Bowler} B.~P.,  et~al., 2010, \mn@doi [\apj] {10.1088/0004-637X/709/1/396},
  \href {https://ui.adsabs.harvard.edu/abs/2010ApJ...709..396B} {709, 396}

\bibitem[\protect\citeauthoryear{Choi, Dotter, Conroy, Cantiello, Paxton  \&
  Johnson}{Choi et~al.}{2016}]{MIST2}
Choi J.,  Dotter A.,  Conroy C.,  Cantiello M.,  Paxton B.,   Johnson B.~D.,
  2016, ApJ, 823, 102

\bibitem[\protect\citeauthoryear{{Clarke}}{{Clarke}}{2009}]{clarke09}
{Clarke} C.~J.,  2009, \mn@doi [\mnras] {10.1111/j.1365-2966.2009.14774.x},
  \href {https://ui.adsabs.harvard.edu/abs/2009MNRAS.396.1066C} {396, 1066}

\bibitem[\protect\citeauthoryear{Deng, Mayer  \& Meru}{Deng
  et~al.}{2017}]{dengetal2017}
Deng H.,  Mayer L.,   Meru F.,  2017, ApJ, 847, 847

\bibitem[\protect\citeauthoryear{{Deng}, {Mayer}, {Latter}, {Hopkins}  \&
  {Bai}}{{Deng} et~al.}{2019}]{hongpingetal19}
{Deng} H.,  {Mayer} L.,  {Latter} H.,  {Hopkins} P.~F.,   {Bai} X.-N.,  2019,
  \mn@doi [\apjs] {10.3847/1538-4365/ab0957}, \href
  {https://ui.adsabs.harvard.edu/abs/2019ApJS..241...26D} {241, 26}

\bibitem[\protect\citeauthoryear{Dotter}{Dotter}{2016}]{MIST1}
Dotter A.,  2016, ApJ, 222, 8

\bibitem[\protect\citeauthoryear{Durisen, Boss, Mayer, Nelson, Quinn  \&
  Rice}{Durisen et~al.}{2007}]{durisenetal07}
Durisen R.~H.,  Boss A.~P.,  Mayer L.,  Nelson A.~F.,  Quinn T.,   Rice W.
  K.~M.,  2007, Gravitational Instabilities in Gaseous Protoplanetary Disks and
  Implications for Giant Planet Formation.
University of Arizona Press

\bibitem[\protect\citeauthoryear{Forgan \& Rice}{Forgan \&
  Rice}{2011}]{forganrice11}
Forgan D.,  Rice K.,  2011, MNRAS, 417, 1928

\bibitem[\protect\citeauthoryear{Forgan \& Rice}{Forgan \&
  Rice}{2013a}]{forganrice13}
Forgan D.,  Rice K.,  2013a, MNRAS, 430, 2082

\bibitem[\protect\citeauthoryear{{Forgan} \& {Rice}}{{Forgan} \&
  {Rice}}{2013b}]{forganrice13b}
{Forgan} D.,  {Rice} K.,  2013b, \mn@doi [\mnras] {10.1093/mnras/stt672}, \href
  {https://ui.adsabs.harvard.edu/abs/2013MNRAS.432.3168F} {432, 3168}

\bibitem[\protect\citeauthoryear{Forgan, Rice, Cossins  \& Lodato}{Forgan
  et~al.}{2011}]{forganriceetal11}
Forgan D.,  Rice K.,  Cossins P.,   Lodato G.,  2011, MNRAS, 410, 994

\bibitem[\protect\citeauthoryear{{Forgan}, {Hall}, {Meru}  \& {Rice}}{{Forgan}
  et~al.}{2018}]{forganetal2018}
{Forgan} D.~H.,  {Hall} C.,  {Meru} F.,   {Rice} W.~K.~M.,  2018, \mn@doi
  [\mnras] {10.1093/mnras/stx2870}, \href
  {https://ui.adsabs.harvard.edu/abs/2018MNRAS.474.5036F} {474, 5036}

\bibitem[\protect\citeauthoryear{Gammie}{Gammie}{2001}]{gammie01}
Gammie C.~F.,  2001, ApJ, 553, 174

\bibitem[\protect\citeauthoryear{Gillon, Triaud, Demory, Jehin, Agol, Deck
  et~al.}{Gillon et~al.}{2017}]{gillonetal17}
Gillon M.,  Triaud A. H. M.~J.,  Demory B.-O.,  Jehin E.,  Agol E.,  Deck
  K.~M.,   et~al., 2017, Nature, 542, 456

\bibitem[\protect\citeauthoryear{Haisch, Lada  \& Lada}{Haisch
  et~al.}{2001}]{haischladalada01}
Haisch K. E.~J.,  Lada E.~A.,   Lada C.~J.,  2001, ApJ, 553, L153

\bibitem[\protect\citeauthoryear{{Hall}, {Forgan}, {Rice}, {Harries},
  {Klaassen}  \& {Biller}}{{Hall} et~al.}{2016}]{halletal2016}
{Hall} C.,  {Forgan} D.,  {Rice} K.,  {Harries} T.~J.,  {Klaassen} P.~D.,
  {Biller} B.,  2016, \mn@doi [\mnras] {10.1093/mnras/stw296}, \href
  {https://ui.adsabs.harvard.edu/abs/2016MNRAS.458..306H} {458, 306}

\bibitem[\protect\citeauthoryear{{Hall}, {Forgan}  \& {Rice}}{{Hall}
  et~al.}{2017}]{halletal2017}
{Hall} C.,  {Forgan} D.,   {Rice} K.,  2017, \mn@doi [\mnras]
  {10.1093/mnras/stx1244}, \href
  {https://ui.adsabs.harvard.edu/abs/2017MNRAS.470.2517H} {470, 2517}

\bibitem[\protect\citeauthoryear{{Hall}, {Rice}, {Dipierro}, {Forgan},
  {Harries}  \& {Alexander}}{{Hall} et~al.}{2018}]{halletal2018}
{Hall} C.,  {Rice} K.,  {Dipierro} G.,  {Forgan} D.,  {Harries} T.,
  {Alexander} R.,  2018, \mn@doi [\mnras] {10.1093/mnras/sty550}, \href
  {https://ui.adsabs.harvard.edu/abs/2018MNRAS.477.1004H} {477, 1004}

\bibitem[\protect\citeauthoryear{{Hall}, {Dong}, {Rice}, {Harries}, {Najita},
  {Alexander}  \& {Brittain}}{{Hall} et~al.}{2019}]{halletal2019}
{Hall} C.,  {Dong} R.,  {Rice} K.,  {Harries} T.~J.,  {Najita} J.,  {Alexander}
  R.,   {Brittain} S.,  2019, \mn@doi [\apj] {10.3847/1538-4357/aafac2}, \href
  {https://ui.adsabs.harvard.edu/abs/2019ApJ...871..228H} {871, 228}

\bibitem[\protect\citeauthoryear{Haworth, Facchini, Clarke  \& Mohanty}{Haworth
  et~al.}{2016}]{haworthetal16}
Haworth T.,  Facchini S.,  Clarke C.,   Mohanty S.,  2016, MNRAS, 475, 5460

\bibitem[\protect\citeauthoryear{{Janson}, {Bonavita}, {Klahr},
  {Lafreni{\`e}re}, {Jayawardhana}  \& {Zinnecker}}{{Janson}
  et~al.}{2011}]{janson11}
{Janson} M.,  {Bonavita} M.,  {Klahr} H.,  {Lafreni{\`e}re} D.,  {Jayawardhana}
  R.,   {Zinnecker} H.,  2011, \mn@doi [\apj] {10.1088/0004-637X/736/2/89},
  \href {https://ui.adsabs.harvard.edu/abs/2011ApJ...736...89J} {736, 89}

\bibitem[\protect\citeauthoryear{{Johnson} et~al.,}{{Johnson}
  et~al.}{2007}]{johnson07}
{Johnson} J.~A.,  et~al., 2007, \mn@doi [\apj] {10.1086/519677}, \href
  {https://ui.adsabs.harvard.edu/abs/2007ApJ...665..785J} {665, 785}

\bibitem[\protect\citeauthoryear{Klee, Illenseer, Jung  \& Duschl}{Klee
  et~al.}{2017}]{kleeetal17}
Klee J.,  Illenseer T.~F.,  Jung M.,   Duschl W.,  2017, A$\&$A, 606, A70

\bibitem[\protect\citeauthoryear{Klee, Illenseer, Jung  \& Duschl}{Klee
  et~al.}{2019}]{kleeetal19}
Klee J.,  Illenseer T.~F.,  Jung M.,   Duschl W.,  2019, A$\&$A, 632, A35

\bibitem[\protect\citeauthoryear{Kratter \& Lodato}{Kratter \&
  Lodato}{2016}]{kratterlodato16}
Kratter K.,  Lodato G.,  2016, Annual Review of Astronomy and Astrophysics, 54,
  271

\bibitem[\protect\citeauthoryear{Kratter \& Matzner}{Kratter \&
  Matzner}{2006}]{krattermatzner06}
Kratter K.,  Matzner C.,  2006, MNRAS, 373, 1563

\bibitem[\protect\citeauthoryear{Kratter, Murray-Clay  \& Youdin}{Kratter
  et~al.}{2010}]{kratteretal10}
Kratter K.,  Murray-Clay R.,   Youdin A.,  2010, ApJ, 710, 1375

\bibitem[\protect\citeauthoryear{Lannier et~al.}{Lannier
  et~al.}{2016}]{lannieretal16}
Lannier J.,  et~al., 2016, A$\&$A, 596, A83

\bibitem[\protect\citeauthoryear{{Laughlin} \& {Bodenheimer}}{{Laughlin} \&
  {Bodenheimer}}{1994}]{laughlinbodenheimer94}
{Laughlin} G.,  {Bodenheimer} P.,  1994, \mn@doi [\apj] {10.1086/174909}, \href
  {https://ui.adsabs.harvard.edu/abs/1994ApJ...436..335L} {436, 335}

\bibitem[\protect\citeauthoryear{Lin \& Pringle}{Lin \&
  Pringle}{1987}]{linpringle87}
Lin D. N.~C.,  Pringle J.,  1987, MNRAS, 225, 607

\bibitem[\protect\citeauthoryear{Lin \& Pringle}{Lin \&
  Pringle}{1990}]{linpringle90}
Lin D. N.~C.,  Pringle J.,  1990, ApJ, 358, 515

\bibitem[\protect\citeauthoryear{Lissauer}{Lissauer}{1993}]{lissauer93}
Lissauer J.,  1993, ARA$\&$A, 31, 129

\bibitem[\protect\citeauthoryear{{Lloyd}}{{Lloyd}}{2011}]{lloyd2011}
{Lloyd} J.~P.,  2011, \mn@doi [\apjl] {10.1088/2041-8205/739/2/L49}, \href
  {https://ui.adsabs.harvard.edu/abs/2011ApJ...739L..49L} {739, L49}

\bibitem[\protect\citeauthoryear{{Lodato} \& {Clarke}}{{Lodato} \&
  {Clarke}}{2011}]{lodatoclarke11}
{Lodato} G.,  {Clarke} C.~J.,  2011, \mn@doi [\mnras]
  {10.1111/j.1365-2966.2011.18344.x}, \href
  {https://ui.adsabs.harvard.edu/abs/2011MNRAS.413.2735L} {413, 2735}

\bibitem[\protect\citeauthoryear{{Lodato} \& {Rice}}{{Lodato} \&
  {Rice}}{2004}]{lodatorice04}
{Lodato} G.,  {Rice} W.~K.~M.,  2004, in {Bertin} G.,  {Farina} D.,   {Pozzoli}
  R.,  eds,  American Institute of Physics Conference Series Vol. 703, Plasmas
  in the Laboratory and in the Universe: New Insights and New Challenges. pp
  266--271 (\mn@eprint {arXiv} {astro-ph/0309570}), \mn@doi{10.1063/1.1718465}

\bibitem[\protect\citeauthoryear{Matzner \& Levin}{Matzner \&
  Levin}{2005}]{matznerlevin05}
Matzner C.,  Levin Y.,  2005, ApJ, 628, 817

\bibitem[\protect\citeauthoryear{{Meru} \& {Bate}}{{Meru} \&
  {Bate}}{2011}]{merubate11}
{Meru} F.,  {Bate} M.~R.,  2011, \mn@doi [\mnras]
  {10.1111/j.1745-3933.2010.00978.x}, \href
  {https://ui.adsabs.harvard.edu/abs/2011MNRAS.411L...1M} {411, L1}

\bibitem[\protect\citeauthoryear{{Meru} \& {Bate}}{{Meru} \&
  {Bate}}{2012}]{merubate12}
{Meru} F.,  {Bate} M.~R.,  2012, \mn@doi [\mnras]
  {10.1111/j.1365-2966.2012.22035.x}, \href
  {https://ui.adsabs.harvard.edu/abs/2012MNRAS.427.2022M} {427, 2022}

\bibitem[\protect\citeauthoryear{Morales et~al.}{Morales
  et~al.}{2019}]{moralesetal19}
Morales J.,  et~al., 2019, Science, 365, 1441

\bibitem[\protect\citeauthoryear{{Nayakshin}}{{Nayakshin}}{2010}]{nayakshin10}
{Nayakshin} S.,  2010, \mn@doi [\mnras] {10.1111/j.1745-3933.2010.00923.x},
  \href {https://ui.adsabs.harvard.edu/abs/2010MNRAS.408L..36N} {408, L36}

\bibitem[\protect\citeauthoryear{{Nielsen} et~al.,}{{Nielsen}
  et~al.}{2013}]{nielsen2013}
{Nielsen} E.~L.,  et~al., 2013, \mn@doi [\apj] {10.1088/0004-637X/776/1/4},
  \href {https://ui.adsabs.harvard.edu/abs/2013ApJ...776....4N} {776, 4}

\bibitem[\protect\citeauthoryear{Nielsen et~al.}{Nielsen
  et~al.}{2019}]{nielsenetal19}
Nielsen E.,  et~al., 2019, ApJ, 158, 13

\bibitem[\protect\citeauthoryear{{Paardekooper}}{{Paardekooper}}{2012}]{paardekooper12}
{Paardekooper} S.-J.,  2012, \mn@doi [\mnras]
  {10.1111/j.1365-2966.2012.20553.x}, \href
  {https://ui.adsabs.harvard.edu/abs/2012MNRAS.421.3286P} {421, 3286}

\bibitem[\protect\citeauthoryear{{Paardekooper}, {Baruteau}  \&
  {Meru}}{{Paardekooper} et~al.}{2011}]{paardekooperetal11}
{Paardekooper} S.-J.,  {Baruteau} C.,   {Meru} F.,  2011, \mn@doi [\mnras]
  {10.1111/j.1745-3933.2011.01099.x}, \href
  {https://ui.adsabs.harvard.edu/abs/2011MNRAS.416L..65P} {416, L65}

\bibitem[\protect\citeauthoryear{{Paczynski}}{{Paczynski}}{1978}]{paczynski78}
{Paczynski} B.,  1978, \actaa, \href
  {https://ui.adsabs.harvard.edu/abs/1978AcA....28...91P} {28, 91}

\bibitem[\protect\citeauthoryear{Pollack, Hubickyj, Bodenheimer, Lissauer,
  Podolak  \& Greenzweig}{Pollack et~al.}{1996}]{pollack96}
Pollack J.,  Hubickyj O.,  Bodenheimer P.,  Lissauer J.,  Podolak M.,
  Greenzweig Y.,  1996, Icarus, 124, 62

\bibitem[\protect\citeauthoryear{Price et~al.}{Price
  et~al.}{2018}]{priceetal18}
Price D.,  et~al., 2018, PASA, 35, e031

\bibitem[\protect\citeauthoryear{{Pringle}}{{Pringle}}{1981}]{pringle81}
{Pringle} J.~E.,  1981, \mn@doi [\araa] {10.1146/annurev.aa.19.090181.001033},
  \href {https://ui.adsabs.harvard.edu/abs/1981ARA%26A..19..137P} {19, 137}

\bibitem[\protect\citeauthoryear{{Rafikov}}{{Rafikov}}{2005}]{rafikov05}
{Rafikov} R.~R.,  2005, \mn@doi [\apjl] {10.1086/428899}, \href
  {https://ui.adsabs.harvard.edu/abs/2005ApJ...621L..69R} {621, L69}

\bibitem[\protect\citeauthoryear{{Rice} \& {Armitage}}{{Rice} \&
  {Armitage}}{2009}]{ricearmitage09}
{Rice} W.~K.~M.,  {Armitage} P.~J.,  2009, \mn@doi [\mnras]
  {10.1111/j.1365-2966.2009.14879.x}, \href
  {https://ui.adsabs.harvard.edu/abs/2009MNRAS.396.2228R} {396, 2228}

\bibitem[\protect\citeauthoryear{Rice, Armitage, Bate  \& Bonnell}{Rice
  et~al.}{2003}]{riceetal03}
Rice W.,  Armitage P.,  Bate M.,   Bonnell I.,  2003, MNRAS, 339, 1025

\bibitem[\protect\citeauthoryear{Rice, Lodato  \& Armitage}{Rice
  et~al.}{2005}]{ricelodatoarmitage05}
Rice W.,  Lodato G.,   Armitage P.,  2005, MNRAS, 364, L56

\bibitem[\protect\citeauthoryear{{Rice}, {Mayo}  \& {Armitage}}{{Rice}
  et~al.}{2010}]{riceetal10}
{Rice} W.~K.~M.,  {Mayo} J.~H.,   {Armitage} P.~J.,  2010, \mn@doi [\mnras]
  {10.1111/j.1365-2966.2009.15992.x}, \href
  {https://ui.adsabs.harvard.edu/abs/2010MNRAS.402.1740R} {402, 1740}

\bibitem[\protect\citeauthoryear{Rice, Armitage, Mamatsashvili, Lodato  \&
  Clarke}{Rice et~al.}{2011}]{riceetal11}
Rice W.,  Armitage P.,  Mamatsashvili G.,  Lodato G.,   Clarke C.,  2011,
  MNRAS, 418, 1356

\bibitem[\protect\citeauthoryear{{Rice}, {Forgan}  \& {Armitage}}{{Rice}
  et~al.}{2012}]{riceetal12}
{Rice} W.~K.~M.,  {Forgan} D.~H.,   {Armitage} P.~J.,  2012, \mn@doi [\mnras]
  {10.1111/j.1365-2966.2011.20153.x}, \href
  {https://ui.adsabs.harvard.edu/abs/2012MNRAS.420.1640R} {420, 1640}

\bibitem[\protect\citeauthoryear{{Rice}, {Paardekooper}, {Forgan}  \&
  {Armitage}}{{Rice} et~al.}{2014}]{riceetal14}
{Rice} W.~K.~M.,  {Paardekooper} S.~J.,  {Forgan} D.~H.,   {Armitage} P.~J.,
  2014, \mn@doi [\mnras] {10.1093/mnras/stt2297}, \href
  {https://ui.adsabs.harvard.edu/abs/2014MNRAS.438.1593R} {438, 1593}

\bibitem[\protect\citeauthoryear{Shakura \& Sunyaev}{Shakura \&
  Sunyaev}{1973}]{shakurasunyaev73}
Shakura N.,  Sunyaev R.,  1973, A$\&$A, 24, 337

\bibitem[\protect\citeauthoryear{{Stamatellos} \& {Whitworth}}{{Stamatellos} \&
  {Whitworth}}{2009}]{stamatellosetal09}
{Stamatellos} D.,  {Whitworth} A.~P.,  2009, \mn@doi [\mnras]
  {10.1111/j.1365-2966.2008.14069.x}, \href
  {https://ui.adsabs.harvard.edu/abs/2009MNRAS.392..413S} {392, 413}

\bibitem[\protect\citeauthoryear{Stamatellos, Whitworth, Bisbas  \&
  Goodwin}{Stamatellos et~al.}{2007}]{stamatellosetal07}
Stamatellos D.,  Whitworth A.,  Bisbas T.,   Goodwin S.,  2007, A$\&$A, 475, 37

\bibitem[\protect\citeauthoryear{Toomre}{Toomre}{1964}]{toomre64}
Toomre A.,  1964, ApJ, 139, 1217

\bibitem[\protect\citeauthoryear{{Vigan} et~al.,}{{Vigan}
  et~al.}{2012}]{vigan2012}
{Vigan} A.,  et~al., 2012, \mn@doi [\aap] {10.1051/0004-6361/201218991}, \href
  {https://ui.adsabs.harvard.edu/abs/2012A&A...544A...9V} {544, A9}

\bibitem[\protect\citeauthoryear{{Vigan} et~al.,}{{Vigan}
  et~al.}{2017}]{viganetal17}
{Vigan} A.,  et~al., 2017, \mn@doi [\aap] {10.1051/0004-6361/201630133}, \href
  {https://ui.adsabs.harvard.edu/abs/2017A&A...603A...3V} {603, A3}

\bibitem[\protect\citeauthoryear{{Young} \& {Clarke}}{{Young} \&
  {Clarke}}{2015}]{youngclarke15}
{Young} M.~D.,  {Clarke} C.~J.,  2015, \mn@doi [\mnras]
  {10.1093/mnras/stv1266}, \href
  {https://ui.adsabs.harvard.edu/abs/2015MNRAS.451.3987Y} {451, 3987}

\bibitem[\protect\citeauthoryear{{Young} \& {Clarke}}{{Young} \&
  {Clarke}}{2016}]{youngclarke16}
{Young} M.~D.,  {Clarke} C.~J.,  2016, \mn@doi [\mnras]
  {10.1093/mnras/stv2378}, \href
  {https://ui.adsabs.harvard.edu/abs/2016MNRAS.455.1438Y} {455, 1438}

\makeatother
\end{thebibliography}

\appendix

\section{Derivation of the Jeans mass in a spiral density perturbation}\label{sec:appendixA}

Deriving the Jeans mass in a spiral density perturbation can be done by equating the freefall timescale of a collapsing spherical density perturbation with the timescale on which the cloud is able to respond to the collapse, given by the sound crossing time.

Starting from the equation of hydrostatic equilibrium for a spherical cloud of density, $\rho (r)$, pressure, $p$, mass, $M(r)$, and radius, $r$, we have,

\begin{equation}
    \frac{dp}{dr} = -\frac{G\rho(r)M(r)}{^2}.
\end{equation}

Newton's second law gives,

\begin{equation}
    \frac{d^2R}{dt^2} = -\frac{GM}{R^2} = -\frac{G}{R^2}\frac{4\pi R_0^2\rho}{3},
\end{equation}

and as,

\begin{equation}
    \frac{d^2R}{dt^2} = \frac{d}{dt}v = \frac{dR}{dt}\frac{dv}{dR} = v\frac{dv}{dR},
\end{equation}

we can state that,

\begin{equation}
    v\frac{dv}{dR} = -\frac{4\pi G R_0^3\rho}{3R^2}.
\end{equation}

Integrating this gives,

\begin{equation}
    \int v dv = -\frac{4\pi GR_0^3\rho}{3}\int\frac{dR}{R^2},
\end{equation}
\begin{equation}
    \frac{1}{2}v^2 = \frac{4\pi GR_0^3\rho}{3R} + C.
\end{equation}

Boundary conditions that $v=0$ when $R=R_0$ give that,

\begin{equation}
    C = -\frac{4\pi GR_0^3\rho}{3},
\end{equation}

so that,

\begin{equation}
    \frac{1}{2}v^2 = \frac{4\pi GR_0^3\rho}{3R} - \frac{4\pi GR_0^2\rho}{3},
\end{equation}
\begin{equation}
    |v| = \sqrt{\frac{8\pi GR_0^2\rho^2}{3}\Big(\frac{R_0}{R} - 1\Big)}.
\end{equation}
To find the timescale on which the cloud will collapse,
\begin{equation}
    t_c = \int dt = \int \frac{dt}{dr}dr = \int \frac{dr}{|v|},
\end{equation}
\begin{equation}
    t_c = \sqrt{\frac{3}{8\pi GR_0^2 \rho^2}}\int _0^{R_0} \Big(\frac{R_0}{R}-1\Big)^{-1/2} dr,
\end{equation}
where,
\begin{equation}
    \int _0^{R_0} \Big(\frac{R_0}{R}-1\Big)^{-1/2} dr \equiv \frac{\pi R_0^2}{2},
\end{equation}
\begin{equation}
    t_c = \sqrt{\frac{3}{8\pi G\rho}}\frac{\pi}{2}=\sqrt{\frac{3\pi}{32G\rho}}.
\end{equation}
The sound crossing time is found by considering the sound speed in the gas cloud, $c_s$, and considering the case of an ideal gas such that,

\begin{equation}
    c_s = \sqrt{\frac{\gamma k_B T}{m}},
\end{equation}
where $k_B$ is the Boltzmann constant, $\gamma$ is the adiabatic index, T is the temperature of the gas, and $m$ is the mass of a gas particle.

The sound crossing time is then,

\begin{equation}
    t_s = \frac{R}{c_s} = \sqrt{\frac{m}{\gamma k_B T}}R.
\end{equation}

When $t_s > t_c$ the cloud will begin to collapse. The Jeans length, $R_J$, is the radius at which the gas cloud will begin to collapse, found when $t_s / t=c \equiv 1$.  

\begin{equation}
    \frac{t_s}{t_c} = \Big(\frac{m}{\gamma k_BT}\Big)^{1/2}R\Big(\frac{32G\rho}{3\pi}\Big)^{1/2},
\end{equation}
\begin{equation*}
    R_J = \sqrt{\frac{3\pi\gamma kT}{32G\rho m}} = c_s\sqrt{\frac{3\pi}{32}\frac{1}{G\rho}}.
\end{equation*}

From this we can then find the Jeans mass as, 

\begin{equation}\label{eq:appendixmj1}
    M_J = \frac{4}{3}\pi R_J^3 \rho_{\rm pert} = \frac{4}{3}\Big(\frac{3}{32}\Big)^{3/2}\pi^{5/2}\frac{c_s^3}{G^{3/2}\rho_{\rm pert}^{1/2}}
\end{equation}

The scale height, \textit{H}, and local surface density of the perturbation, $\Sigma_{\rm pert}$ are related to $\rho_{\rm pert}$ as,

\begin{equation}\label{eq:appendixrhopert}
    \rho_{\rm pert} = \Sigma_{\rm pert}/2H,
\end{equation}

where,

\begin{equation}\label{eq:appendixsigma}
    \Sigma_{\rm pert} = \Big(1 + \frac{\Delta\Sigma}{\Sigma}\Big).
\end{equation}

Here, $\frac{\Delta\Sigma}{\Sigma}$ represents the fractional amplitude of the spiral wave perturbation.

Rearranging equation \ref{eq:Q} in terms of G and $\Sigma$ gives,

\begin{equation}\label{eq:appendixtoomre}
    (G\Sigma)^{1/2} = \Big(\frac{c_s \Omega}{\pi Q}\Big)^{1/2}.
\end{equation}

Now substituting \ref{eq:appendixsigma}, \ref{eq:appendixtoomre} and $H = c_s/\Omega$ into \ref{eq:appendixmj1} gives,

\begin{equation}\label{eq:appendixmj2}
    M_J = \frac{4\sqrt{2}}{3}\Big(\frac{3}{32}\Big)^{3/2}\frac{\pi^3}{G}\frac{Q^{1/2}c_s^2 H}{(1+\frac{\Delta\Sigma}{\Sigma})^{1/2}}
\end{equation}

In the presence of external irradiation \cite{riceetal11} showed that,

\begin{equation}
    \langle \frac{\Sigma_{\rm RMS}}{\Sigma} \rangle = 4.47 \sqrt{\alpha}.
\end{equation}

Substituting this into \ref{eq:appendixmj2} gives the expression for Jeans mass found in equation \ref{eq:jeansmass},

\begin{equation}\label{eq:appendixmj3}
    M_J = \frac{\sqrt{3}}{32G}\frac{\pi^3Q^{1/2}{c_s}^2 H}{(1+4.47\sqrt{\alpha})^{1/2}}.
\end{equation}


\bsp	
\label{lastpage}
\end{document}